\documentclass[twocolumn,showpacs,preprintnumbers,amsmath,amssymb]{revtex4}
\usepackage{graphicx}
\usepackage{dcolumn}
\usepackage{bm}

\begin{document}


\title{Generalization of the Second Law for a Nonequilibrium Initial State}

\author{H.-H. Hasegawa$^1$}
\altaffiliation[]{Center for Complex Quantum Systems, Univ. of Texas, Austin, Texas, 78712 USA}
\email{hhh@mx.ibaraki.ac.jp}
\author{J. Ishikawa$^1$} 
\author{K. Takara$^1$}
\author{D.J. Driebe$^2$}
\affiliation{
$^1$Department of Mathematical Sciences, Ibaraki University\\
Bunkyo, Mito, 310-8512 Japan\\
$^2$Division of Math, Science and Technology, Nova Southeastern University,
Fort Lauderdale, Florida, 33314 USA
}


\date{\today}

\begin{abstract}
We generalize the second law of thermodynamics in its maximum work formulation for a nonequilibrium initial distribution. It is found that in an isothermal process, the Boltzmann relative entropy (H-function) is not just a Lyapunov function but also tells us the maximum work that may be gained from a nonequilibrium initial state. The generalized second law also gives a fundamental relation between work and information. It is valid even for a small Hamiltonian system not in contact with a heat reservoir but with an effective temperature determined by the isentropic condition. Our relation can be tested in the Szilard engine, which will be realized in the laboratory.
\end{abstract}



\maketitle 

The maximum work formulation of the second law of thermodynamics  
relates the work needed to move a system from one equilibrium state to  
another to the free energy difference between those states.  It tells  
us that the work must be greater than or equal to the  
difference in free energies. In recent refinements of the second law, such as the fluctuation theorem\cite{Evans1993} and the Jarzynski equality\cite{Jarzynski1997}\cite{Crooks1999}, the maximum work formulation in the sense of the average work is rigorously shown for an initial canonical distribution\cite{Jarzynski1997}.

In this letter we generalize the maximum work formulation of the second law to transitions between nonequilibrium states. Our generalization allows one to find the maximum work that can be gained from such a transition and the processes that realize it. We are able to consider isolated systems as well as those coupled to a heat reservoir and both adiabatic and isothermal transitions.

The derivation is based on the Jarzynski equality modified for a  
nonequilibrium initial distribution.  This leads to a relation between  
the work and the Boltzmann relative entropy with an effective  
temperature.  The Boltzmann relative entropy, also known as the  
Kullback--Leibler divergence, is always positive and gives a  
``distance'' between the nonequilibrium initial distribution and the  
canonical distribution\cite{Information}.

For a finite Hamiltonian system without a heat reservoir, the effective temperature is determined by an isentropic condition. The maximum work is realized in two successive processes: an instantaneous stabilization of the nonequilibrium initial distribution and an isentropic process.

When the system is coupled to a large heat reservoir, the effective temperature is the temperature of the heat reservoir. From the generalized second law, the maximum work is realized in two successive processes: an instantaneous stabilization and a quasi-static isothermal process.
For an isolated system the maximum work is realized by an initial stabilization followed by an isentropic process.

The generalized second law gives a fundamental relation between work and information like the Landauer principle\cite{Landauer}. Actually, we are able to derive the Landauer principle by applying our generalized maximum work formulation to the Szilard engine\cite{Szilard}\cite{Bennett}\cite{Shizume}. We will also derive the Sagawa--Ueda generalization of the second law\cite{Sagawa--Ueda} with the Maxwell demon\cite{Maxwell} from our generalization.  Our formulation would play an important role in an experimental realization of informational cooling as first proposed by Leo Szilard\cite{Szilard}\cite{MarkRaizen}.

In this letter, we consider a Hamiltonian system. The total system is divided into two parts, a small system and a large heat reservoir. 
$X=(x,y)$ is a phase space point of the total system where  $x$ ($y$) is a point of the systemireservoir). 

The total Hamiltonian is a sum of three Hamiltonians, the system, 
$H^{(S)}(x,a(t))$, the reservoir, $H^{(R)}(y)$, and the  
interaction between them, $H^{(I)}(x,y)$, as
\begin{equation}
H(X,a(t))
=H^{(S)}(x,a(t))+H^{(R)}(y)+H^{(I)}(x,y)
\end{equation}
where $a(t)$ is a time-dependent parameter associated with external operations.  We operate on or observe only the system and not the reservoir. We assume no transition of particles (variables) between the system and the reservoir. Hereafter, we represent a function in the system with the superscript, $(S)$. 

The external operations are given by changing the external parameter, $a(t)$, following a given protocol. For convenience, we assume that the external parameter is constant within unit intervals and instantaneously changes at integer times. That is, $a(t)=a_{\tau}$ for\\
$t\in [\tau,\tau+1)$~$(\tau=0,1,2,...,T)$. Hereafter, we abbreviate 
$H_\tau(X)=H(X,a_{\tau})$ for convenience. 

A nonequilibrium distribution, $\rho_0(X)$, is prepared at $t=0$.
The time evolution of the probability distribution of the total system is governed by the Hamiltonian dynamics as
\begin{equation}
\rho_{\tau}(X)=
U_{\tau-1}U_{\tau-2}\cdot\cdot\cdot U_{0}\rho_0(X)
\end{equation}
where $U_{\tau}=\exp(-{\rm i L}_{\tau})$ is the time evolution operator and ${\rm L}_{\tau}$ is the Liouvillian corresponding to the Hamiltonian $H_\tau$.

The canonical distribution for the Hamiltonian $H_\tau(X)$ is defined as 
\begin{equation}
\rho_{{\rm can},\tau}(X,\alpha)
=\exp\{\alpha(F_{\tau}(\alpha)-H_{\tau}(X))\}
\end{equation}
where $\alpha$ is a parameter like an inverse temperature and the free energy is   
$F_{\tau}(\alpha)=-\log\{<1|\exp(-\alpha H_{\tau})>\}/\alpha$
where $<A|B>\equiv\int A^*(X)B(X) dX$ in the bra-ket notation.
The canonical distribution is invariant, 
$U_{\tau}\rho_{{\rm can},\tau}=\rho_{{\rm can},\tau}$, for any $\alpha$, since it is a function of $H_{\tau}(X)$.

The work is given as the sum of contributions from T phase points at $t=\tau$ $(\tau=1,2,...,T)$ on a trajectory starting at $X_0$,
$\{X_{\tau}(X_0)\}$,  
\begin{equation}
W(X_0)=\sum _{\tau=1}^{T} \{H_{\tau}(X_{\tau}(X_0))
-H_{\tau-1}(X_{\tau}(X_0))\}.
\end{equation}
The expectation value of the work is given as
\begin{equation}
<W>\equiv<W|\rho_0>.
\end{equation}

It is important to note that ``the maximum work'' done by the system is  the lowest work done on the system in our definition. Hereafter we will discuss a lower bound of the expectation value of the work.

We generalize the maximum work formulation for a nonequilibrium initial
distribution by starting with the Jarzynski equality\cite{Jarzynski1997},
\begin{equation}
<\exp\{\alpha(\Delta F(\alpha)-W)\}|\rho_{{\rm can},0}(\alpha)>=1
\end{equation}
where $\Delta F(\alpha)=F_{T}(\alpha)-F_{0}(\alpha)$. For an invertible  
nonequilibrium initial distribution, we can modify it as 
\begin{equation}
<\exp\{\alpha(\Delta F(\alpha)- W)-\log(\frac{\rho_0}{\rho_{{\rm can},0} 
(\alpha)})\}|\rho_0>=1
\end{equation}
By applying the Jensen inequality, $<{\rm e}^{-x}>~\geq{\rm e}^{-<x>}$, 
we obtain the following important inequality\cite{Kawai},
\begin{equation}
<W>~\geq \Delta F(\alpha) 
- \frac{1}{\alpha}{\rm D}(\rho_0|\rho_{{\rm can},0}(\alpha))
\end{equation}
where $D(\rho_A|\rho_B)=<\log(\rho_A)-\log(\rho_B)|\rho_A>$ is the non- 
negative Kullback--Leibler divergence, also known as the Boltzmann relative entropy, which gives the ``distance'' between two probability distributions, $\rho_A$ and $\rho_B$.
We note that it is valid for any $\alpha$, since the the Jarzynski equality is based on the invariance of the canonical distribution.

We determine an effective temperature  
$\tilde{\beta}^{-1}$ as the $\alpha^{-1}$, which maximizes the lower bound, 

\noindent${\rm Max}_{\alpha}[\Delta F(\alpha) 
- \frac{1}{\alpha}{\rm D}(\rho_0|\rho_{{\rm can},0}(\alpha))]$.
The maximization condition requires the following isentropic property,
\begin{equation}
<\log(\rho_{{\rm can},T}(\tilde{\beta}))|\rho_{{\rm can},T}(\tilde{\beta})>
=<\log(\rho_0)|\rho_0>
\end{equation}

We first consider an isolated Hamiltonian system without the reservoir.
Then, $H_{\tau}=H^{(S)}_{\tau}$ and $\rho_0=\rho^{(S)}_0$. The generalized maximum work formulation for a nonequilibrium initial distribution in an adiabatic process is obtained with the effective temperature,
\begin{equation}
<W>~\geq~ <H^{(S)}_{T}|\rho^{(S)}_{{\rm can},T}(\tilde{\beta})>
-<H^{(S)}_0|\rho^{(S)}_0>.
\end{equation}

We show by what processes the maximum work is realized in the isolated Hamiltonian system without the reservoir. We formally introduce a Hamiltonian, ${\cal H}^{(S)}_0$, for which the nonequilibrium initial distribution can be written as a canonical one, 
$\rho^{(S)}_0
=\exp(\tilde{\beta}({\cal F}^{(S)}_0(\tilde{\beta})-{\cal H}^{(S)}_0))$ 
where 
${\cal F}^{(S)}_0(\tilde{\beta})$ is the free energy for ${\cal H}^{(S)}_0$.
Then, Eq.(10) is rewritten as 
\begin{equation}
<W>~\geq~ <W>_{\rm IS}+<W>_{\rm IE}
\end{equation}
where
\begin{eqnarray}
<W>_{\rm IS}&\equiv&
<{\cal H}^{(S)}_{0}-H^{(S)}_0|\rho^{(S)}_0>~~~~~~~~~~~~~~~~~~~~~~~~~\\ 
<W>_{\rm IE}&\equiv&<H^{(S)}_T|\rho^{(S)}_{{\rm can},T}(\tilde{\beta})>
-<{\cal H}^{(S)}_{0}|\rho^{(S)}_0>.
\end{eqnarray}
The most efficient way to gain work from the nonequilibrium  
initial distribution in an adiabatic process is as follows: (1) Change the initial Hamiltonian, $H^{(S)}_0$, to ${\cal H}^{(S)}_{0}$ for instantaneous stabilization (IS) of the initial distribution. (2) Change 
${\cal H}^{(S)}_{0}$ to the final one, $H^{(S)}_T$ in an isentropic process (IE). Here we assumed $T$ is large enough and the isentropic process from 
$\rho^{(S)}_0$ to $\rho^{(S)}_{{\rm can},T}(\tilde{\beta})$ exists.

Now we couple the system to the reservoir. We prepare the initial distribution separated from the reservoir at $t=0$. Then we put the system in contact with the reservoir for intermediate times. Finally, we take the system away from the reservoir at $t=T$.  We assume localized distributions for the both system and reservoir and a short range interaction. 

We assume that the reservoir is in equilibrium with temperature $\beta^{-1}$at $t=0$. Then, the total initial probability distribution is given as 
\begin{equation}
\rho_{0}=\exp\{\beta(F^{(R)}(\beta)-H^{(R)}(y))\}
\rho^{(S)}_{0}(x)
\end{equation}
where $F^{(R)}(\beta)$ is the free energy of the reservoir.

The effective temperature is determined by the maximization of the right hand side of Eq.(8), after substituting Eq.(14) into Eq.(8). After careful estimation of the cancellation of two large quantities in the Kullback-- Leibler divergence, the effective temperature becomes the temperature of the reservoir. In the estimation, we assumed the energy of the system and the work are negligible compared with the energy of the reservoir.

The generalized maximum work formulation for a nonequilibrium initial distribution in an isothermal process is obtained as
\begin{equation}
<W>~\geq \Delta F^{(S)}(\beta)-\frac{1}{\beta}
{\rm D}(\rho^{(S)}_0|\rho^{(S)}_{{\rm can},0}(\beta))
\end{equation}
where we neglect the interaction at the both $t=0$ and $t=T$. The Boltzmann relative entropy (H-function), with the temperature of the reservoir, has not only meaning as a Lyapunov function but also tells us the maximum work that may be gained from the nonequilibrium initial state. 

We show by what processes the maximum work is realized in the Hamiltonian system with the reservoir. We can rewrite Eq.(15) as
\begin{equation}
<W>~\geq~<W>_{\rm IS}+<W>_{\rm QI}
\end{equation}
where
\begin{eqnarray}
<W>_{\rm IS}&\equiv&
 <{\cal H}^{(S)}_{0}-H^{(S)}_0|\rho^{(S)}_0>\\ 
<W>_{\rm QI}&\equiv&
F^{(S)}_T(\beta) - {\cal F}^{(S)}_0(\beta) 
\end{eqnarray}
where 
$\rho^{(S)}_0=\exp\{\beta({\cal F}^{(S)}_0(\beta)-{\cal H}^{(S)}_0)\}$.
Similar to the case of an adiabatic process, the most efficient way to gain work in an isothermal process is as follows: (1) Change the initial Hamiltonian, $H^{(S)}_0$, to ${\cal H}^{(S)}_{0}$ for the instantaneous stabilization (IS). (2) Change ${\cal H}^{(S)}_{0}$ to the final one, 
$H^{(S)}_T$, in a quasi-static isothermal process (QI). Here we assumed $T$ is large enough and the quasi-static isothermal process from $\rho^{(S)}_0$ to $\rho^{(S)}_{{\rm can},T}(\beta)$ exists.

The Clausius relation in an isothermal process is also generalized. From the first law of thermodynamics, The absorbed heat is given as
\begin{equation}
<Q>~\equiv~-<W>+<H_T|\rho^{(S)}_T>-<H_0|\rho^{(S)}_0>
\end{equation}
where $\rho^{(S)}_T(x)\equiv\int dy \rho_T(x,y)$.
Then, the maximum absorbed heat is obtained from the maximum work as
\begin{eqnarray}
<Q_{\rm max}>~&\equiv&~-\beta^{-1}
<\log(\rho^{(S)}_{{\rm can},T}(\beta))|\rho^{(S)}_{{\rm can},T}(\beta)>
\nonumber\\
& &~+\beta^{-1}<\log(\rho^{(S)}_0)|\rho^{(S)}_0>
\end{eqnarray}
In the generalized Clausius relation, the maximum absorbed heat in an isothermal process is given as the difference between the final thermodynamic entropy and the initial Shannon entropy\cite{MaesTasaki2007}.

Here we mention a further generalization of the maximum work formulation for a certain set of trajectories (paths for a quantum system), which start from a nonequilibrium initial distribution and reach a final observable (detector). It is important for experiments where only particles entering a detector are observed. This can be modeled by choosing the final observable, $A(X_T)$, which projects out a nonequilibrium final state. This generalization gives us the maximum work for a transition between two nonequilibrium states. The generalization is straightforward using 
$<A|\rho_{{\rm can},T}>$ instead of 
$<1|\rho_{{\rm can},T}>$ in derivation of the modified Jarzynski equality.

The generalized maximum work formulation of thermodynamics for a nonequilibrium initial distribution gives an important relation between two important concepts in physics, the energy and the information. It is also a generalization of the Landauer principle\cite{Landauer}. When we apply the generalized maximum work formulation to the Szilard engine\cite{Szilard}\cite{Shizume}, the famous relation,
\begin{equation}
<W>~\geq -\frac{1}{\beta}\log(2),
\end{equation}
is immediately obtained, since the height of the one sided initial distribution, which has one bit of information, is just twice of the right-left symmetric canonical one in the double-well potential.

Recently Sagawa and Ueda have generalized the Landauer principle for quantum measurement\cite{Sagawa--Ueda}. They showed how the maximum work is changed by measuring a quantum system as the Maxwell demon\cite{Bennett}. One of their important results is that the maximum work is given as the Shannon entropy with temperature in the most efficient classical measurement. We will derive it from our generalized second law to demonstrate its validity.

Suppose the density matrix of a quantum system is the canonical one with the temperature, $\beta^{-1}$, before the measurement. It is given as 
\begin{equation}
P_{{\rm can}}=\sum_n r_n(\beta) R_n
\end{equation}
where $R_n$ is the projection operator for the $n$th energy level, $E_n$, and $r_n(\beta)=\exp\{\beta(F(\beta)- E_n)\}$ is the probability to measure the system at the $n$th energy level.

After the most efficient classical measurement, the system may be in the 
$n$th energy level, $P_n=R_n$, with the probability $r_n(\beta)$.
From our generalized second law, in an isothermal process  
the work is bounded as
\begin{equation}
W_n\geq -\frac{1}{\beta}{\rm Tr}[\{\log(P_n)-\log(P_{\rm can}(\beta))\}P_n]
=\frac{1}{\beta}\log(r_n(\beta))
\end{equation}
As Sagawa and Ueda established, the expectation value is given as the Shannon entropy,
\begin{equation}
<W>~\geq \frac{1}{\beta}\sum_n r_n(\beta)\log(r_n(\beta)).
\end{equation}

It is interesting to apply this argument for an adiabatic process.
The effective temperature, $\tilde{\beta}^{-1}_n$, is determined by the following isentropic condition,
\begin{equation}
{\rm Tr}[\log(P_{\rm can}(\tilde{\beta}_n))P_{\rm can}(\tilde{\beta}_n)]
={\rm Tr}[\log(P_n)P_n].
\end{equation}
The effective temperature becomes zero in the most efficient classical measurement. From our generalized maximum work formulation, in an adiabatic process the work from the $P_n$ is bounded as
\begin{equation}
W_n\geq 
\sum_i\lim_{\tilde{\beta}_i\rightarrow\infty}r_i(\tilde{\beta}_i)E_i-E_n
=E_0-E_n.
\end{equation}
As we expected, we can gain all energy in the most efficient classical measurement. The expectation value of the work is given as
\begin{equation}
<W>~\geq E_0 - \sum_n r_n(\beta) E_n.
\end{equation}
The maximum work in an isothermal process is greater than the one of the corresponding adiabatic process without the reservoir. This is clear from the definition of the effective temperature. In the adiabatic process, we can gain only the energy of the system without heat from the reservoir.
For both adiabatic and isothermal processes the maximum work decreases in a non-efficient measurement\cite{separatingpaper}.

The generalized second law we introduce in this letter gives a relation
between energy and information that is applicable even outside of
thermodynamics.  In such contexts the thermodynamic temperature is
replaced by an effective temperature determined by an isentropic
condition.  In this way the second law becomes universal parallel to
the first law, which is the universal law of energy conservation.  As Ilya
Prigogine asserted, the second law plays an important role not only
in large thermodynamic systems but in small systems, such as elementary
particles.

Our generalized second law is valid for any Markov process with the invariant canonical distribution. Therefore, it is applicable to a system that may be modeled as a stochastic process, such as the Langevin equation in an isothermal process. It is interesting to apply our maximum work formulation to a nonequilibrium steady state studied in such stochastic systems\cite{HatanoSasa2001} and make clear  from this perspective how house-keeping heat appears. We would also consider a Carnot-like cycle operation by combining adiabatic processes in Hamiltonian dynamics and isothermal processes in stochastic models.

The results given in this letter open new perspectives for addressing long-standing issues in the study of irreversible processes. Among those issues are the role of dynamical instability such as chaos, the reconciliation of reversible Hamiltonian dynamics with the approach to equilibrium, and the definition of nonequilibrium thermodynamic entropy in a reversible Hamiltonian dynamics. These will be addressed in a forthcoming paper that will consider the generalized maximum work formulation in a chaotic Hamiltonian system\cite{forthcomingpaper}. The important properties of Hamiltonian dynamics such as time-reversal symmetry, the Liouville law, and the invariance of the Shannon entropy, which we did not consider in this letter, will be discussed there.

\begin{acknowledgments}
H.H.H. and D.J.D. acknowledge the late Professor Ilya Prigogine for his encouragement, suggestions, and support of this work. H.H.H., J.I., and K.T. thank N. Okamoto for his contributions. This work was supported by a Grant-in-Aid for Scientific Research (Grant No.17540346) from the Ministry of Education, Culture, Sports, Science, and Technology of Japan.
\end{acknowledgments}

\bibliographystyle{plain}

\end{document}